\documentclass[9pt,twocolumn,twoside,lineno]{article}
\usepackage[utf8]{inputenc}
\usepackage[T1]{fontenc}
\usepackage{graphicx} 
\usepackage{hyperref}
\usepackage{lineno,color}
\usepackage{amsmath,amsfonts,amssymb}
\usepackage{authblk}
\usepackage{geometry}

\title{Geography of Science: Competitiveness and Inequality}

\author[a,e,1]{Aurelio Patelli}
\author[b,$\dagger$]{Lorenzo Napolitano}
\author[c,a,e]{Giulio Cimini}
\author[d,a,e]{Andrea Gabrielli}

\affil[a]{\footnotesize Enrico Fermi Center for Study and Research, via Panisperna 89a, 00184, Rome (Italy)}
\affil[b]{European Commission, Joint Research Center (JRC), 41092 Seville (Spain)}
\affil[c]{Physics Department and INFN, University of Rome Tor Vergata, 00133 Rome (Italy)}
\affil[d]{Engineering Department, University ``Roma Tre'', 00146 Rome (Italy)}
\affil[e]{Institute for Complex Systems (CNR), UoS Sapienza, 00185 Rome (Italy)}

\date{}

\begin{document}

\maketitle

\begin{abstract}
	Using ideas and tools of complexity science we design a holistic measure of \textit{Scientific Fitness}, encompassing the scientific knowledge, capabilities and competitiveness of a research system.
	We characterize the temporal dynamics of Scientific Fitness and R\&D expenditures at the geographical scale of nations, highlighting patterns of similar research systems, and showing how developing nations (China in particular) are quickly catching up the developed ones. 
	Down-scaling the aggregation level of the analysis, we find that even developed nations show a considerable level of inequality in the Scientific Fitness of their internal regions. 
	Further, we assess comparatively how the competitiveness of each geographic region is distributed over the spectrum of research sectors. 
	Overall, the Scientific Fitness represents the first high quality estimation of the scientific strength of nations and regions, opening new policy-making applications for better allocating resources, filling inequality gaps and ultimately promoting innovation.
\end{abstract}
\normalsize


\section{Introduction}
Science is based on the progressive augmentation of existing knowledge building on past discoveries, through a recursive process involving empirical observation and the formulation of testable hypotheses. 
Similarly to what happens for technological innovation and economic growth~\cite{Hidalgo2009,Tacchella2013,Pugliese2017}, scientific progress requires appropriate capabilities: previous knowledge, tools, human capital, resources, and so on. The combination and interaction of such capabilities, even from different contexts, pushes the boundary of science through new knowledge and discoveries, as well as through re-discoveries via previously uncharted paths~\cite{Dosi2000,Tria2014,Iacopini2020}.
This process naturally occur mostly in geographic areas where many different capabilities are concentrated~\cite{Balland2020}, whence we can assume that the scientific output of a region reflects the set of relevant capabilities available.

The quantitative evaluation of scientific outcomes, from the microscopic level of individual researchers and institutions to the macroscopic case of entire nations, is nowadays a common practice~\cite{Waltman2016,Fortunato2018}. 
At the macro level, a seminal work by May~\cite{May1997} assessed the performance of national research systems using an index borrowed from the economic literature: the Revealed Comparative Advantage (RCA)~\cite{Balassa1965}, computed on the number of scientific documents produced by each nation in the various research sectors. 
King~\cite{King2004} pursued a different approach, ranking nations according to the share of global citations received by their document output, and introduced funding as an additional variable of the analysis. 
Subsequently, the use of citations became the gold standard for assessing research quality, and several metrics with swinging performance have been proposed -- see~\cite{Waltman2016} for a comprehensive review of the field. 
However this approach has recently been questioned, due for example to the very different amount of resources that nations invest in scientific research. 
In fact, even for the most economically developed nations, the scientific success measured on citations and the public spending in research and development (R\&D)~\cite{Cimini2016,laverde2019can} (as well as returns to innovation~\cite{Patelli2017}) are correlated but also present strong deviations, and therefore should be considered as complementary dimensions for a correct evaluation of scientific performance. 
Another important problem is given by the presence of bias and distortions in citation patterns~\cite{Abramo2014, Abramo2016}. 
Indeed the dynamics of the citation process strongly depend on sector-specific characteristics, and citation statistics are often distorted by the presence of outliers (the few documents attracting a huge amount of citations)~\cite{aksnes2004effect,althouse2009differences}. These and other issues may reduce the explanatory power of citation-based metrics, as well as their variants based on top-percentage citations~\cite{Brito2018}, including the \textit{H-index}~\cite{Montazerian2019}.

There are two additional key aspects that citation share metrics do not take into account.
On the macroscopic scale, nations do not specialize in a few research sectors but tend to diversify their activity into as many sectors as possible. 
This is explained by the capability scheme, for which a given geographic area is active in all research sectors allowed by the capabilities that are present on its territory. 
Since capabilities are heterogeneously distributed, nations have a heterogeneous level of diversification, thus diversification itself can be used as a basic proxy of scientific performance. 
In addition, while nations with many different capabilities (typically, the developed economies) are competitive in almost all existing research sectors, nations with fewer capabilities (the less developed economies) perform well only in a few research areas with a lower degree of sophistication or complexity.
Such a \textit{nested} structure, induced by the capability scheme, indicates the presence of a competitive mechanism shaping the connections amidst the scientific actors -- akin to what is observed in natural ecosystems~\cite{Mariani2019} as well as in human productive activities~\cite{Hidalgo2009}.  
Indeed, although the scientific environment is neither directly nor indirectly aimed at the production of physical goods or services (for which there is a clear payoff) and is not subject to the incentives of competitive markets, there are actually many sources of competition, since most research systems rely on merit-based processes to determine funding, hiring, careers, and thus indirectly scientific research itself. Therefore, only naively science can be considered as guided by non-competitive actors who collaborate for the advancement of knowledge.

Overall, the nested pattern observed when comparing national research systems~\cite{Cimini2014} suggests that diversification and composition of the scientific research basket can be used to measure the scientific competitiveness (or Fitness) of a nation; at the same time, the complexity of a research sector depends on its ubiquity and on the Scientific Fitness of nations that are competitive in that sector. 
The Economic Fitness and Complexity (EFC)~\cite{Tacchella2012,Cristelli2013,Cristelli2015} algorithm is the ideal tool to estimate the fixed point of this circular relation. 
The purpose of this work is precisely to develop a framework for quantifying scientific competitiveness by leveraging the EFC toolbox. 

In a nutshell, we build an appropriate database for our analysis starting from the Open Academic Graph (OAG)\cite{OAG}~\cite{MAG1,MAG2,MAKG}, a freely accessible collection of information about individual scientific publications, covering a large portion of the scientific production corpus. 
On the one hand, OAG assigns documents to geographic areas according to the location of the research institutes to which the authors are affiliated. On the other hand, OAG assigns documents to research sectors according to a hierarchical classification of scientific topics, each known as \textit{Field of Studies} (FoS). 
The documents produced by a geographical area in a research sector provide a basic measure of scientific performance through an appropriate count of citations received. In this analysis we can use a variable resolution both in terms of geographical scale (we follow the Territorial Level scheme implemented by the OECD~\cite{oecd-geo}) and of FoS hierarchical level. 
Filtering this data using the RCA allows obtaining the scientific bipartite network (SBN hereafter) connecting geographic areas with the research sectors they are competitive in, and finally computing the Scientific Fitness of such areas through the EFC algorithm~\cite{Cimini2014}. 
Note that our approach follows the path initially outlined by May~\cite{May1997}, though we compute RCA not on document production but on citation counts, in accordance with the ideas proposed by King~\cite{King2004} --- and likewise we complement the analysis with data about monetary resources invested in scientific research. 
In particular we use Higher Education expenditures on Research \& Development (HERD), again provided by the OECD~\cite{OECD_HERD}. 
We refer the reader to the Materials and Methods section for a more detailed description of the workflow.

\section{Results}

\begin{figure*}
	\centering
	\includegraphics[scale=0.8]{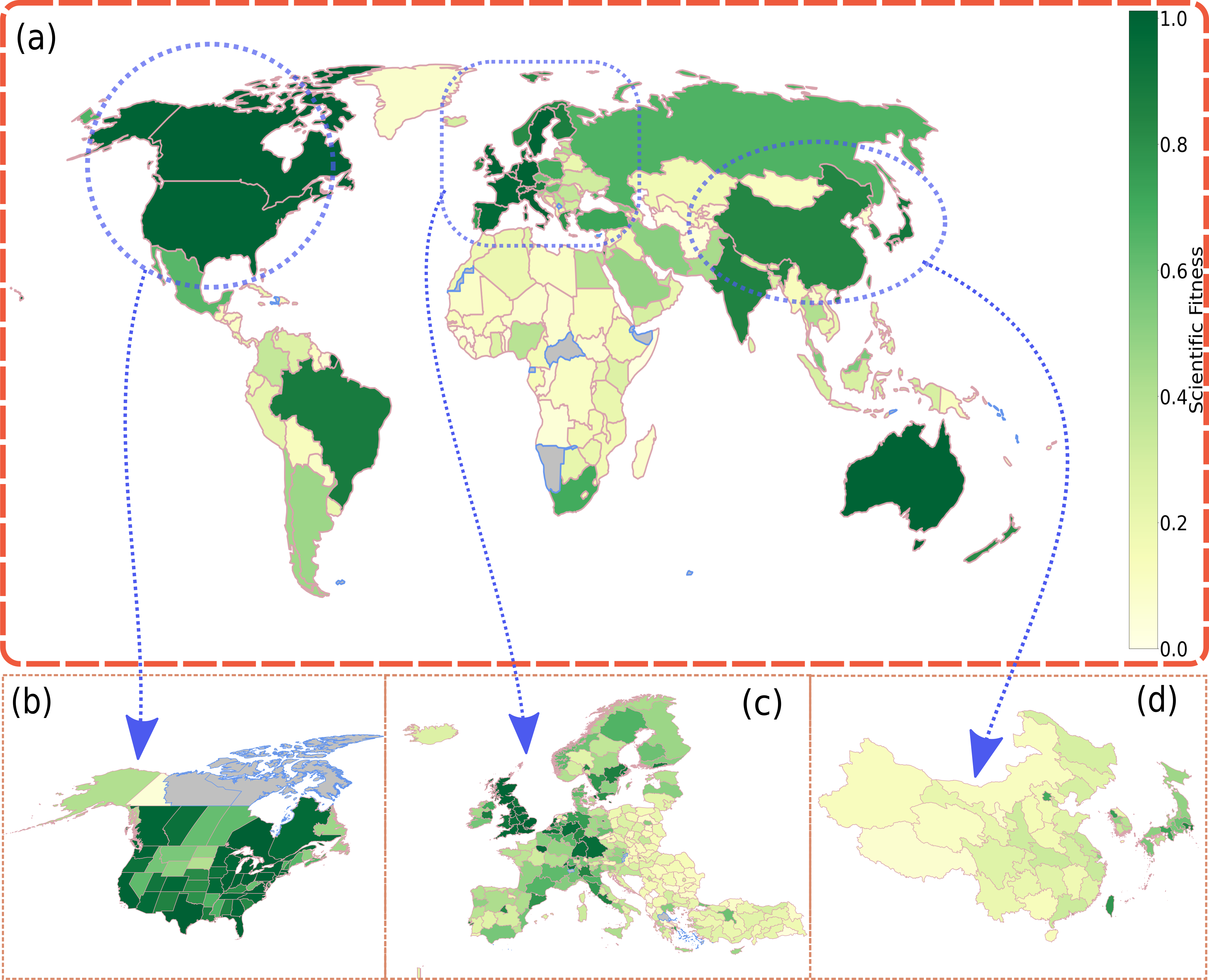}
	\caption{Map of the Scientific Fitness of nations (TL1, panel a) and of regions (TL2) within North America (panel b), Europe and Turkey (panel c) and China, Japan and South Korea (panel d).
		The color scale indicates the average Fitness between 1998 and 2018 (missing entries are colored in gray), with darker and lighter tone for higher and lower Fitness, respectively (the scale $[0,1]$ is the same for the national and regional levels). 
		Notice how the Fitness of a nation cannot be simply obtained by summing nor averaging the Fitness of its regions (see Figure~\ref{fig:tl2_results} below).
		The elliptic projection of the map follows the Robinson projection (esri:54030).
	}
	\label{fig:maps}
\end{figure*}


We start by discussing Scientific Fitness at the geographic scale of nations -- corresponding to Territorial Level 1 (TL1) of the OECD classification. 
Unsurprisingly the geographical distribution of Fitness values, reported in the top map of Figure \ref{fig:maps}, shows that the most developed and rich nations are also the top performers in science, while the developing nations are ranked lower~\cite{Cimini2014}. 
Such heterogeneous patterns are similar to those associated with traditional measures of economic size or relevance (such as GDP or population) but have a more intensive character, since small countries can display a high Scientific Fitness while large ones may not, \textit{e.g.} Switzerland ranks higher than India. 
A higher correlation is observed with the Economic Fitness computed using export data~\cite{Tacchella2012,Cristelli2013,Cristelli2015}, which is also aimed to measure competitiveness based on owned capabilities, tough there is not a one-to-one correspondence between the two measures~\cite{JRC_report} (see SI for a further comparison). 
Notably, the raking of Scientific Fitness is also different from that obtained using metrics based on citation shares, such as the Mean Normalized Citation Score (MNCS)~\cite{Waltman2016,Cimini2016}, which measure research efficiency rather than competitiveness.
Indeed, MNCS ranks at the top the small but efficient research systems -- such as Switzerland, Israel and Singapore. Instead Scientific Fitness accounts both for efficiency (through the use of the RCA filter) and diversification (i.e.,  the cumulative stock of capabilities owned by a nation), and thus allows for a more fair comparison between small and large research systems.
Remarkably the same patterns are observed also when the analysis is performed using a different dataset (we report in the Supporting Information the case of Scimago~\cite{ScimagoDatabase}, based on Scopus).

\begin{figure*}
	\centering
	\includegraphics[scale=0.275]{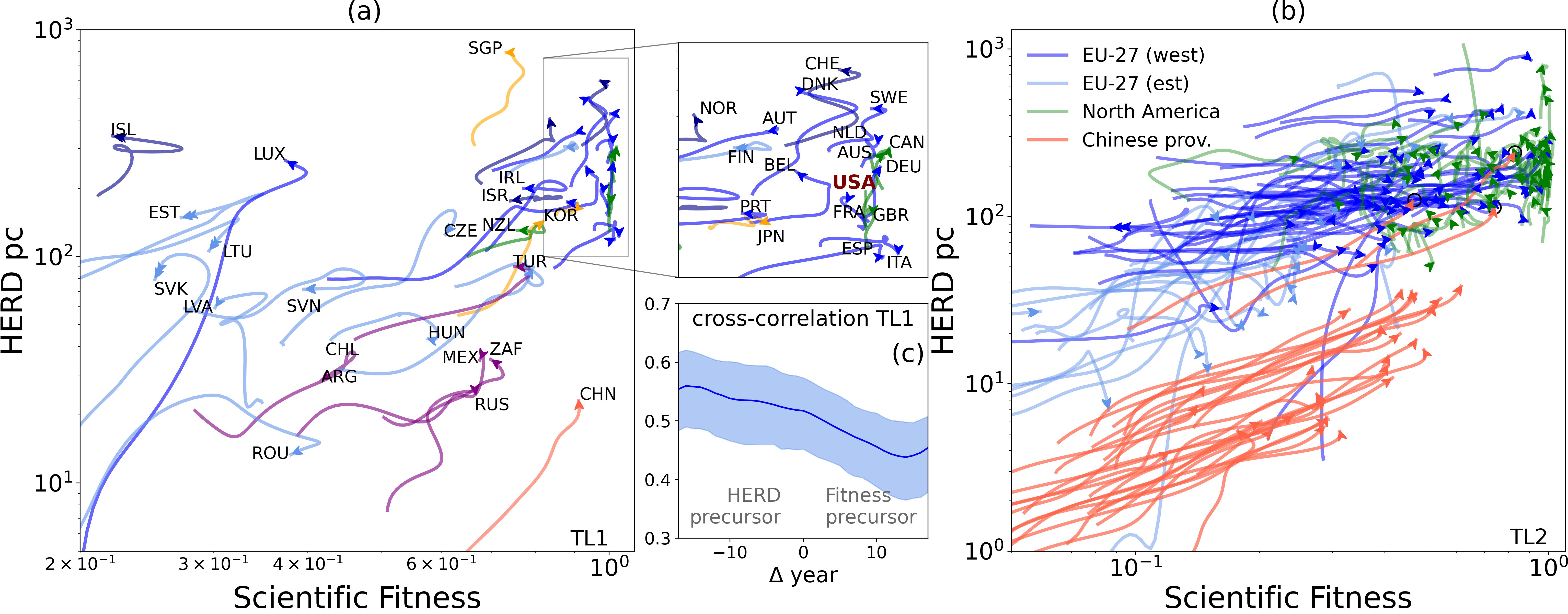}
	\caption{(panel a) Trajectories of nations (TL1) in the plane defined by Scientific Fitness and resources invested, the latter  measured by Higher Education expenditures on R\&D per capita (HERD-pc). Line colors are used to group nations into macro-areas: dark blue for west EU nations (plus Switzerland, Israel, Norway, Island), light blue for east EU nations, , green for the English-speaking nations (United States, United Kingdom, Canada, Australia, New Zealand) red for China, yellow for the other Asian nations (Singapore, South Korea, Japan) and purple for middle-income countries (Russia, South Africa, Mexico, Argentina, Chile). Trajectories represent data from 2000 to 2017, with the arrow indicating the direction of time. 
		The inset zooms on the top-right corner where there is a concentration of highly competitive nations. 
		(panel b) Trajectories are also displayed for regions (TL2) belonging to China and a selection of EU west, EU east and North America nations. 
		(panel c) Cross-correlation between Scientific Fitness and HERD at the national scale (TL1) averaged over the whole set of countries as a function of the temporal delay ($\Delta$ year) used to compute these quantities. The blue contour represents the $25-75\%$ quantile, generated with a bootstrapping technique. Note that a cross-correlation value of about 0.5 is comparable to analogous estimations carried out in the economics context~\cite{Patelli2021}
	}
	\label{fig:expenditure_diagram}
\end{figure*}

Following previous literature on Science of Science~\cite{King2004,Abramo2014,Cimini2016}, we obtain a richer picture by complementing Scientific Fitness with the amount of resources that are invested in scientific research. 
A similar approach (with some caveats discussed below) is also used in the classic EFC literature, where Economic Fitness is scattered against a monetary measure of income (typically the Gross Domestic Product per capita); the dynamics in the two dimensional space defined by these variables highlight clusters of similar economies, allowing for a very precise economic forecasting~\cite{Tacchella2018}.
As already mentioned, here we employ Higher Education expenditures on R\&D (HERD)~\cite{OECD_HERD}, namely the expenditures for basic research performed in the higher education sector, which among the sources of public funding are those most connected to scientific performance as measured through citations of published documents~\footnote{The other sources of public funding are~\cite{OECD_HERD}: the Business Expenditures on R\&D (BERD), namely R\&D expenditures performed in the business sector, which is mostly related to the creation of new products and production techniques (patents); the Government Intramural Expenditures on R\&D (GOVERD), namely expenditures in the government sector, which is often mission-oriented and therefore less connected to publication outputs (see~\cite{Cimini2016} and the discussion therein). 
	In the Supporting Information we show results of analysis performed using Gross Expenditure on R\&D (GERD), given by the sum of HERD, BERD and GOVERD.}. 
This data is available only for OECD members and a few other important economies (such as China and Russia); therefore the following analysis will be focused only on this subset of high and middle income countries. 
Figure~\ref{fig:expenditure_diagram} in panel-(a) shows the trajectories of these nations in the two dimensional plane defined by Scientific Fitness and HERD per capita (HERD-pc). 
We observe that the most developed economies usually concentrate in the top right corner of the diagram (enlarged in the inset) characterized by high values of both Fitness and HERD-pc. 
The other nations are instead scattered along the diagonal, for which Scientific Fitness is proportional to resources invested, and their trajectories are typically directed towards the top-right region: these countries are quickly catching up with the most advanced ones.
Off-diagonal trajectories provide interesting information, similar to those obtained within the economic framework~\cite{Cristelli2013}. 
The top left corner contains small national research systems with peculiar features, where investments are not efficiently turned into scientific competitiveness. 
This is for instance the case of Iceland, which does not attract much attention in terms of citations, and of Luxembourg, where the presence several private firms headquarters may bias the scientific production to patent-related documents~\cite{Patelli2017}. 
In the opposite corner, China (and to a minor extent Russia, South Africa and Mexico) features a high scientific competitiveness despite low public R\&D expenditures, with both quantities growing quickly in time.

\begin{figure*}
	\centering
	\includegraphics[scale=0.3]{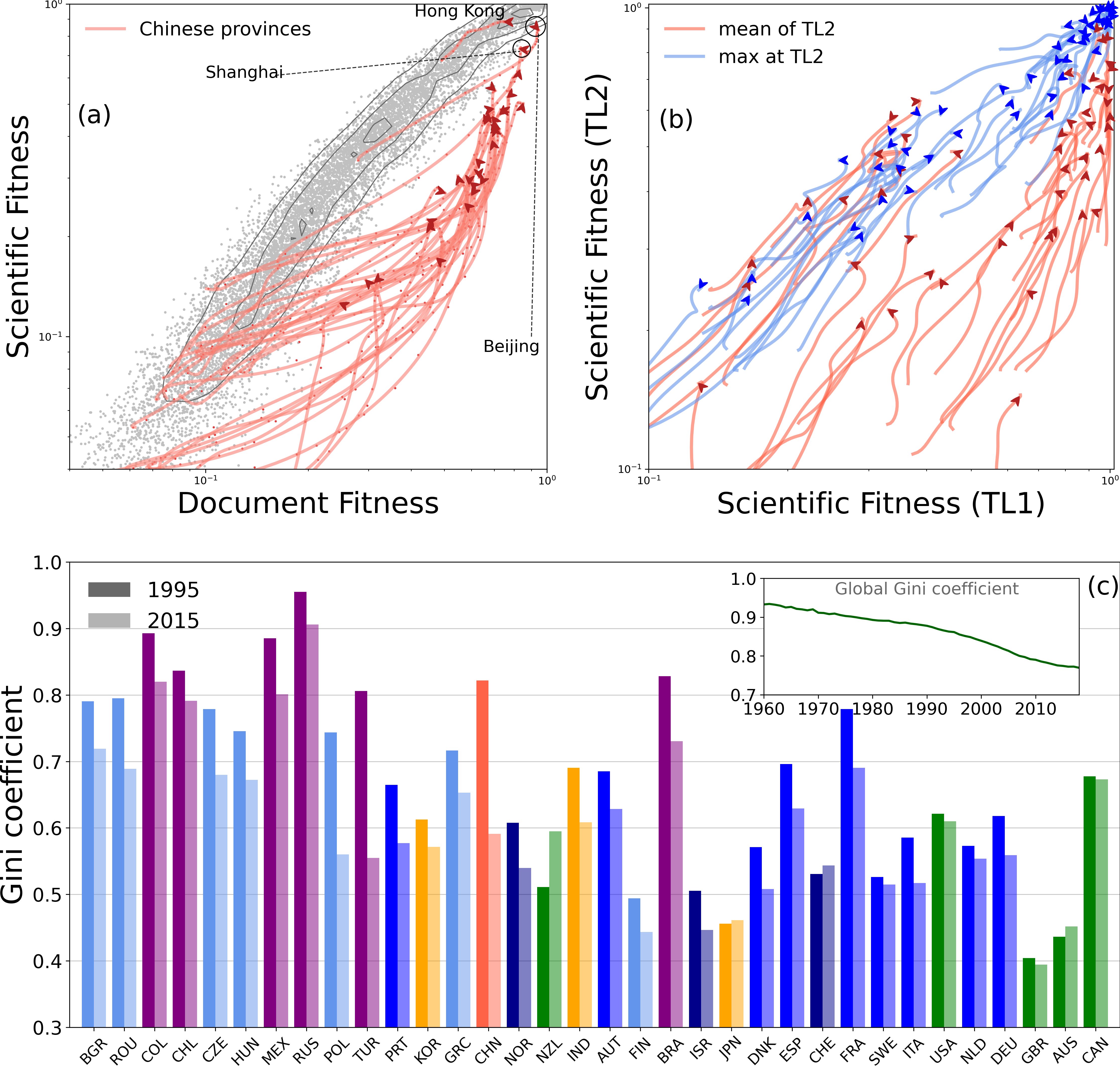}
	\caption{(panel a) Comparison of Scientific Fitness and document Fitness (i.e., Fitness computed using published documents) at the regional level (TL2). The black lines indicate the density level contour of the cloud of points while each red trajectory indicates the evolution of a Chinese province. The trajectories map the evolution from 2000 to 2018, with the arrow indicating the direction of time.
		(panel b) Comparison between the Scientific Fitness of nations, computed either at the national level (TL1) or as the mean (red line) or maximum (blue line) of the Fitness of internal regions (TL2).
		(panel c) Gini coefficients of each nation, computed over the citation counts of internal regions. We report values for two years: 1995 (full color bars) and 2015 (shade color bars). Nations are ordered according to their average Scientific Fitness in the central decade (2000-2010). The inset represents the temporal evolution of the Gini coefficient of the whole world.}
	\label{fig:tl2_results}
\end{figure*}

A main advantage of our framework is the possibility to perform the analysis of Scientific Fitness at a more detailed geographic level, in order to highlight the competitiveness of specific regions within nations. 
The bottom maps of Figure \ref{fig:maps} report the Scientific Fitness of regions (as defined by Territorial Level 2 (TL2) of the OECD classification) for three macro-areas: North America, Europe and East Asia. 
We observe a recurrent pattern for which the Fitness of a nation is mostly concentrated in its capital region (also because capitals typically host the headquarters of the largest national research institutions). 
The English-speaking nations (United States and United Kingdom, and the same happens for Australia) are the exception by featuring high Fitness in all their regions. 
Such a widespread competitiveness can be also due to the language bias of the dataset, which covers non-English literature only partially, especially for Social Sciences and Humanities~\cite{Sivertsen2012} (see Materials and Methods and further analysis in the Supporting Information), and possibly to the advantage of native English speakers in better writing scientific articles which therefore attract more citations.
The evolution of Scientific Fitness and HERD-pc at the regional level is shown in the right panel of Figure~\ref{fig:expenditure_diagram}.
We see again that while most of the North Americans regions are top performers, Fitness values of European regions form a cloud ranging from low to high competitiveness. 
The case of China stands alone: only three provinces (Beijing, Hong Kong and Tianjin) belong to the cloud of EU regions, while the others follow a very regular flow with a steady increase both in competitiveness and public expenditures. 
Indeed China invested enormously in science starting from the end of the last century, with growing expenditures in R\&D throughout the country. 
Apart from the three outliers, the competitiveness of Chinese provinces has not yet reached that of the western countries regions, but it will eventually do~\cite{Huang2018,Xie2019}. 
This can be clearly seen in panel-(b) of Figure~\ref{fig:tl2_results}, where the trajectories of regional Scientific Fitness are scattered with those of \textit{document Fitness}, i.e., competitiveness computed on document production rather than citation accrued (see the Materials and Methods). 
Mainland Chinese provinces follow a unique pattern. Their document Fitness has increased substantially in the considered time span (2000-2018), due to growing resources and the consequent acquisition of new capabilities. However, initially this research output was not able to capture many citations from the international scientific community, likely due to a low initial level of competitiveness.
Only recently Chinese research became very competitive and started to attract citations, with a consequent growth in Scientific Fitness: the trajectories of Chinese provinces are quickly moving towards the main cluster where the regions of other countries are located.

\begin{figure*}
	\centering
	\includegraphics[scale=0.175]{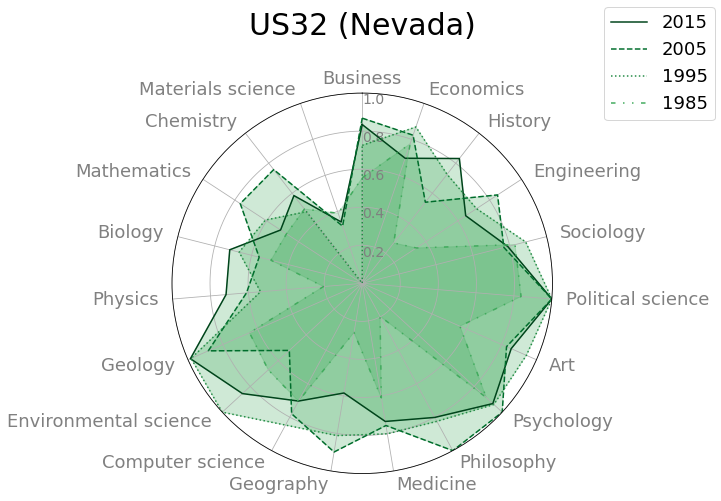}
	\includegraphics[scale=0.175]{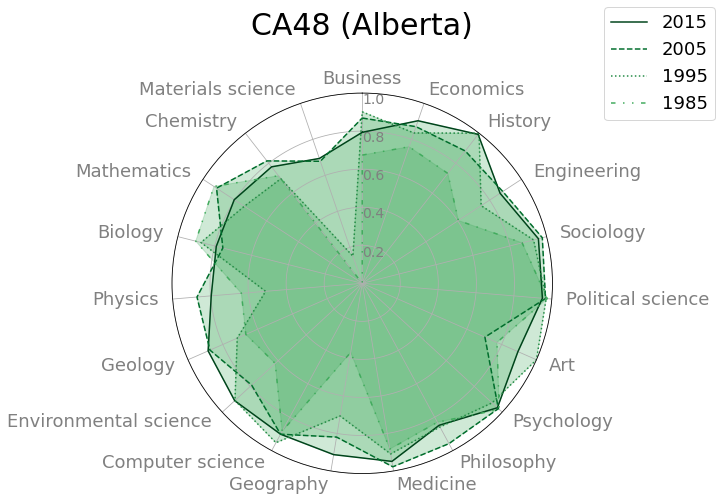}
	\includegraphics[scale=0.175]{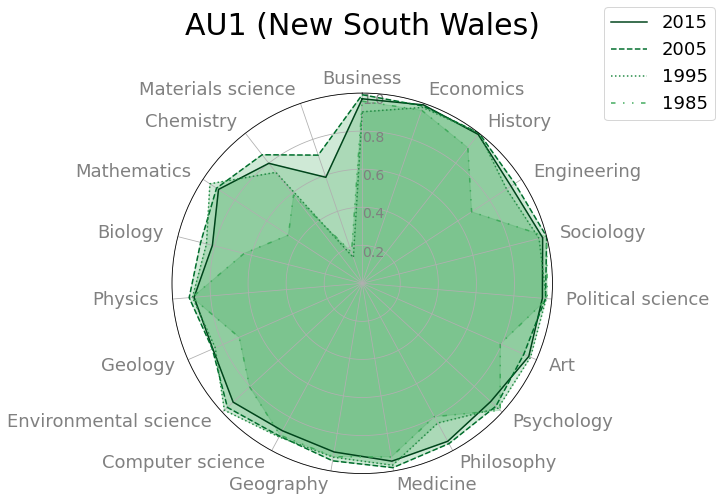}
	\includegraphics[scale=0.175]{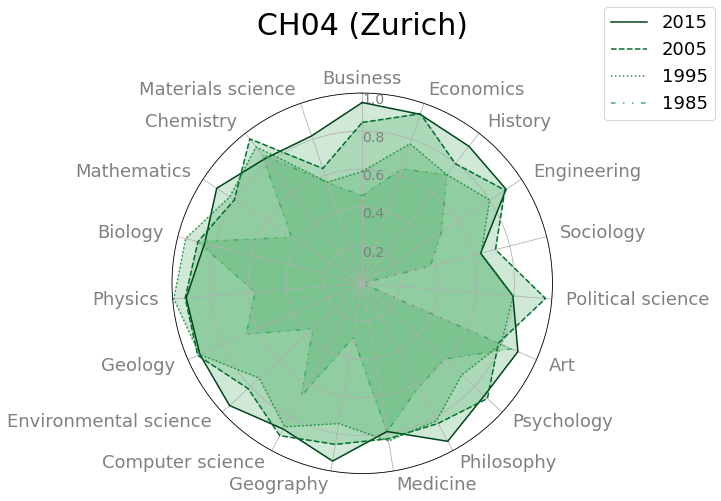}
	\includegraphics[scale=0.175]{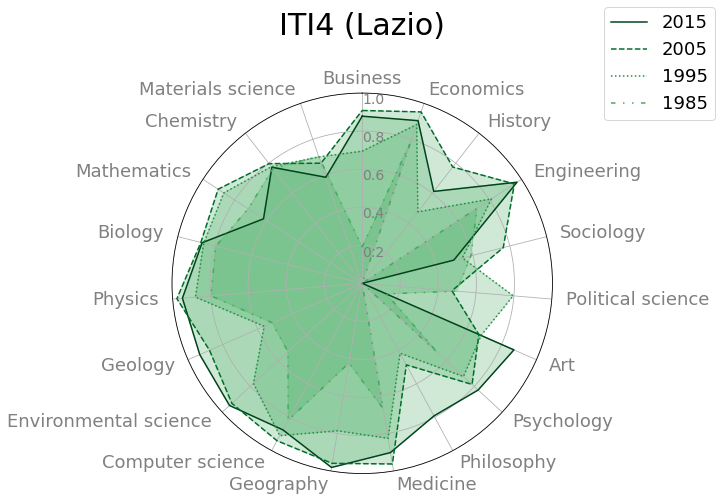}
	\includegraphics[scale=0.175]{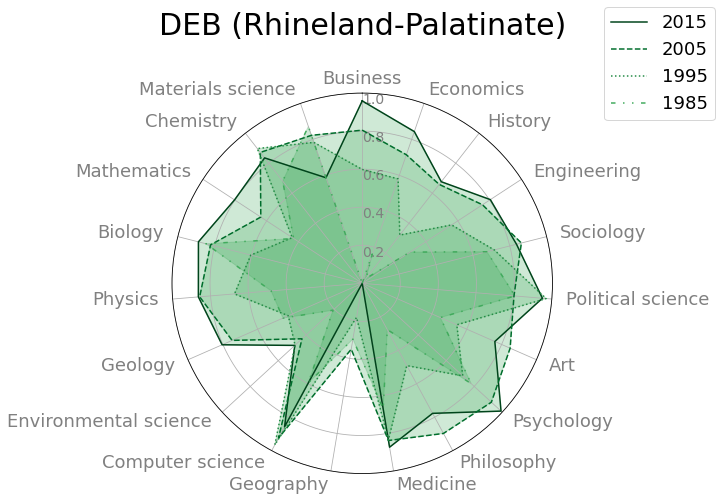}
	\includegraphics[scale=0.175]{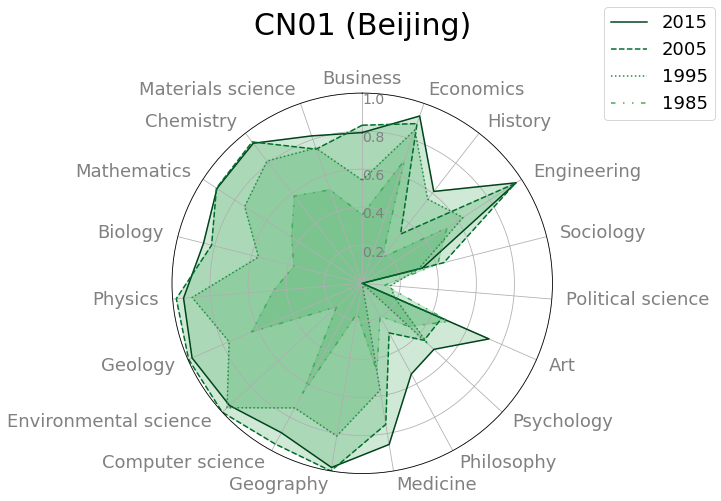}
	\includegraphics[scale=0.175]{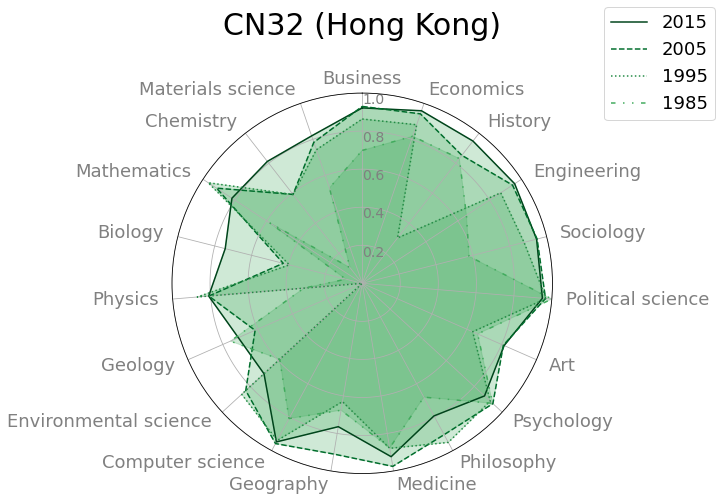}
	\includegraphics[scale=0.175]{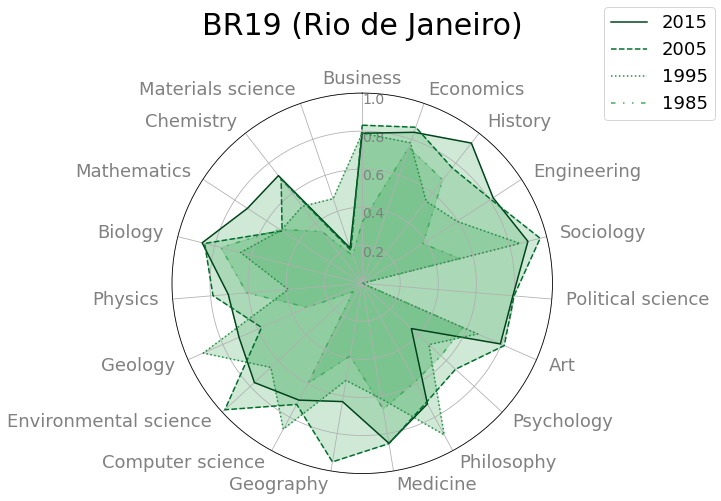}
	\caption{Radar plots of the scientific sector Fitness of different sample regions (TL2). 
		Top row: Nevada (USA), Alberta (Canada), New South Wales (Sydney). 	Central row: Zurich (Switzerland), Lazio (Italy), Rhineland-Palatinate (Germany). 
		Bottom row: Beijing (China), Hong Kong (China), Rio de Janeiro (Brazil).
		Sectors are ordered clockwise with decreasing average complexity (Business is the most complex and Material Science is the less complex sector). The radar lines indicate how Fitness has evolved over the course of thirty years, from 1985 to 2015. 
	}
	\label{fig:sector_Fitness}
\end{figure*}

Overall the results of the analysis at TL2 indicate that the Fitness of a nation is not obtained by simply averaging or summing up the Fitness of its regions, because the most exclusive capabilities are typically concentrated only in a few regions, which thus determine the  national competitiveness. 
This is confirmed by the plot in panel-(a) of Figure~\ref{fig:tl2_results}, which shows that the national Fitness is more correlated to the Fitness of its most competitive region (Pearson correlation of about 0.96) rather than to the mean Fitness of its regions (Pearson correlation of about 0.70). 
More importantly, our framework highlights the strong heterogeneity of Fitness values both across and within nations, and thus allows locating the geographical inequalities of the scientific research system.
Values of the Gini coefficients (see Materials and Methods for the precise mathematical definition of the Gini coefficients implemented) are shown in the bottom  panel of Figure~\ref{fig:tl2_results} for the available nations with more than 4 TL2 regions, and for two reference years (1995 and 2015) spaced by two decades. 
The analysis shows that the United Kingdom and Australia have the lower inequality score and in general the English-speaking nations feature low inequalities, while mid-income countries are characterized by the highest inequality levels. 
We also compute the global Gini coefficient over all available regions in the world; the Inset in panel-(c) of Figure~\ref{fig:tl2_results} shows that the global level of inequality is slowly decreasing in time. 

Down-scaling the analysis from nations (TL1) to regions (TL2) means increasing the geographical resolution of our method. Similarly, we can increase the resolution regarding the research sectors, by exploiting the hierarchical classification of FoS. 
Thus, for example, instead of computing the total Scientific Fitness of a geographic area we can compute its \textit{sector} Fitness restricted to one of the 19 entries of the FoS main hierarchical level. 
Figure~\ref{fig:sector_Fitness} shows the radar plots of the sector Fitness for some example regions. The 19 research sectors are ordered clockwise in the radar according to their complexity (computed as the average complexity of their sub-sectors), so that Business is the most complex and Material Science is the less complex FoS.
Note that the EFC algorithm typically assigns higher complexity to soft sciences (Economics, Social Sciences and Humanities) rather than to medical and hard sciences, because it turns out that only the most competitive players are active in the former sectors, while the latter sectors are more ubiquitous. 
This pattern can be partially due to the aforementioned bias of our bibliometric data towards English-speaking nations in soft sciences. 
However a more fundamental explanation exists: only the most developed research systems have reached the level of capabilities required to perform scientific research in, e.g., Business Administration, Environmental Ethics and Cognitive Science. 
These sectors require solid prerequisites in the hard sciences, but they are not necessarily related to high technological requirements\footnote{Note that the average complexity of a research sector does not fully reflect the complexity of the associated sub-sectors. Indeed also in the hard sciences there are highly sophisticated research sectors that require expensive instruments and large collaborations. For example, while the average complexity of \textit{Business} is 1.82, the complexity of \textit{Polymer science}, a child code of \textit{Material science} and \textit{Chemistry}, is as high as 5.34.}, rather they are aimed at addressing the most advanced needs of a society~\cite{Cimini2014}. 
Overall, the analysis of the scientific sector Fitness allows to quantitatively detect the strengths and weaknesses of each region, as well as their temporal evolution. 
For instance Figure~\ref{fig:sector_Fitness} shows how the Beijing region experienced a fast growth in competitiveness in the hard sciences while it still falls back in artistic and cultural areas with respect to western regions. Regions like Zurich, Lazio and Alberta have instead a more uniform pattern of competitiveness, especially in the last decades. Note how the top-performing regions like New South Wales can also have competitive gaps but only in the less complex sectors.

\section{Discussions and Conclusion}
This work aims to bring together two recent lines of research: \textit{Science of Science}~\cite{Waltman2016,Zeng2017,Fortunato2018}, which develops quantitative methods and assessment tools to study the evolution of science itself, and \textit{Economic Fitness and Complexity}~\cite{Hidalgo2009,Tacchella2012,Hidalgo2021}, which aims at measuring the productive capabilities of economic systems. Indeed, our framework to assess competitiveness in scientific research builds on the theory of hidden capabilities and employs properly calibrated bibliometric indicators. The proposed methods allow for a consistent comparison between different geographical areas and research sectors at varying level of resolution. In this work we presented only a handful of applications, highlighting the heterogeneity of scientific competitiveness among nations as well as the inequalities within national research systems. We further characterized the performance of scientific actors across the various research sectors, and showed that the evolution of research systems can be properly described using two dimensions, Scientific Fitness and R\&D expenditure. In the plane defined by these variables, nations form clusters of similar research systems operating within countries that have reached comparable stages of development.

Similarly to other the classic applications in the EFC literature, this study shows that a high explanatory and forecasting power is achieved when Economic Fitness is coupled with a variable related to the amount of resources available in the system under enquiry. Typically, the EFC literature proxies resource endowments with Gross Domestic Product (GDP)~\cite{Tacchella2018}; for our purposes, HERD is is the more appropriate measure. 
However there is a fundamental difference between the use of GDP and HERD.
GDP is a measure of generated capital and wealth, hence it reflects the outcome of the production process; for this reason GDP can be interpreted as a a consequence of Economic Fitness.
Instead, HERD measures the amount of public resources that are fed into the scientific system and thus is an input requisite for Scientific Fitness.
Consequently, while both the trajectories of countries in the GDP-Fitness plane and the trajectories in the Scientific Fitness-HERD allow to extract interesting patterns concerning the way in which nations cluster in the plane, there are also remarkable differences in their interpretation. 
For instance, there is evidence suggesting that one can qualitatively predict 5-years trends in GDP in the light of the historical evolution of economic fitness~\cite{Tacchella2018}. 
However, it would be overconfident to push the analogy between Economic Fitness and Scientific Fitness to the point of trying to infer future Scientific Fitness from historical HERD data. 
A comprehensive analysis of the relation between Scientific Fitness and different measures of input and output of research systems represents a promising avenue for future research.

In addition to uncovering non-trivial patterns in the evolution of national and regional knowledge production systems, the application of the EFC methodology to the realm of scientific production data also has the potential relevance for policy making.
Even though the direct concern of Economic policy is not so much knowledge creation, but rather Economic output or innovation, it is known that competitiveness in scientific fields is robustly linked to the development of competitive advantages in patenting as well as export~\cite{Pugliese2017}.
Since success in one of the above three layers -- knowledge, innovation, trade -- tends to be a precursor of success in the others, it is reasonable to argue that a long-sighted approach to growth and development policies can only benefit from factoring knowledge production capabilities into the equation.
Finally, the analysis of the scientific competitiveness of regional areas add a tool in the analysis of local capabilities, necessary in the developments of less wealthy regions.


\section{Material and Methods}
We extract the scientific database from the Open Academic Graph (OAG) (\url{https://www.microsoft.com/en-us/research/project/open-academic-graph/}), a freely available snapshot of a two billion-scale academic graph resulting from the unification of Microsoft Academic Graph and AMiner~\cite{MAG1,MAG2,MAKG}. We use OAG v2, created at the end of November 2018. The database is composed by a list of entries related to various scientific literature: journal articles, books, conference proceedings, reviews, and others. The OAG coverage is estimated to be comparable to that of Scopus or Wos~\cite{hug2017coverage}, thus likely presenting similar geographical and phonetic biases -- in particular the partial coverage of non-English written literature, especially in the Social Sciences and Humanities where research output is often published in the native language~\cite{Sivertsen2012}.
The OAG data spans more than a century, starting in principle at the beginning of 1800. In practice, data before the Second World War presents large fluctuations mainly due to the scarce amount of scientific production for most of the regions.
Hence, we start the analysis in 1960, although the core results are presented only for the recent decades where also expenditure data is available (see below).

The classification of research sectors is defined by the Fields of Study (FoS), features which are dynamically evaluated by an ``\textit{in-house knowledge base related entity relationship, which is calculated based on the entity contents, hyperlinks and web-click signals}''~\cite{MAG1}.
The FoS are mostly organized into a hierarchical structure, with the main characteristic that a code may have more parents~\footnote{The very few exceptions of codes that are labelled at a fine level but without information on their parents are removed from the analysis. This does not represent a problem, since we consider only the highest levels of the FoS hierarchy.}. 
This structure presents a static layer 0 with 19 hand-defined codes, corresponding to the main classification of the research sectors. Moving deeper in the hierarchy, layer 1 presents 294 codes while layer 2 has more than 80 000 codes and this number may change in time when new FoS are generated~\footnote{Deeper layers 3, 4 and 5 mostly split the larger topics, but are not considered in the present work.}.


The OAG database is used to construct the bipartite network linking geographical areas to research sectors.
To this end we select only the OAG entries with full information on authors' affiliation, FoS, citations count and year of publication. 
Using this data we build tables reporting, for each year, the number of scientific documents produced by each geographical areas in the various FoS, and their citations received up to the OAG creation date. In order to assign a document to a geographic area, OAG uses the location of the authors' main affiliation. 
Note that in some cases it is not possible to select a precise location because the affiliation may address generically to a multinational firm or a multi-location research council (such as CNRS in France or CNR in Italy). 
In these cases the location of the headquarter is used, although this process may artificially boost capital regions. 
Note also that there are several documents labeled by multiple FoS and/or with several author affiliations. 
In these cases we employ a fractional counting approach, by assigning the document to FoS and geographic areas with a weight that is inversely proportional to number of FoS and number of authors~\footnote{For example, if a paper is labeled with FoS $s_1$ and $s_2$ and has three authors, the first two affiliated with (also different) institutions in area $g_1$ and the third with an institution in area $g_2$, the paper is assigned to FoS $s_1$ and $s_2$ with the same weight 1/2, while it is assigned to geographic areas $g_1$ and $g_2$ respectively with weights $2/3$ and $1/3$. The paper's citations are split according to the same ratios.}.
Fractional counting has the main advantage that allows aggregating tables on both the geographical and FoS dimensions without increasing disproportionately the weight of the most productive actors. 
Additionally, fractional counting has to be preferred as it better balances the scientific outputs of large and small geographical regions~\cite{Aksnes2012}. 

Following the classical approach of the \textit{Scientometrics} literature, we use citations received by scientific documents as a reliable proxy for the quality of research~\cite{Waltman2016}. 
However, the simple citations count presents a few drawbacks, especially when used to assess a small corpus of papers. This is due to the time papers need to reach a stable level of citations~\cite{Burrell2002}, to the high skewness of the citation distribution for single papers~\cite{Radicchi2008,Romeo2003}, and to the dependence of citation patterns on the specific sector and journal considered. 
Indeed the dynamical process underlying the evolution of citation counts is well modeled using a preferential attachment process~\cite{Medo2011,Eom2011,Wang2013}.
This means that the sum of the citations accrued by a set of papers is dominated by the citations of the few most cited outliers, which are in turn subject to strong statistical fluctuations (especially in small sets). 
A simple yet effective approach to reduce such fluctuations as well as the skewness of the citations distribution consists in using a logarithmic transformation~\cite{Fairclough2015,Medo2016}.
Thus we employ the \textit{log-citations} count
\begin{eqnarray}
	w_{gs} = \log (1+c_{gs})
	\label{eq:logcitation}
\end{eqnarray}
where $g$ labels a geographical area and $s$ a research sector, while $c_{gs}$ is the citation count of documents assigned to area $g$ for FoS $s$ published in a given year.

We further filter the log-citations counts to build a Scientific Bipartite Network (SBN), relating for each year the geographical areas on one set with the research sectors in which they are competitive on the other set. To this end we use an index borrowed from the economics literature, the Revealed Comparative Advantage (RCA)~\cite{Balassa1965}, which measures competitiveness as the ability of an actor to perform an activity more that a reference level -- the latter given by the global average performance of the selected activity. Applied to our case study, a geographical area is considered competitive in a research sector if its RCA is above a threshold, typically set to 1. 
In formula:
\begin{equation}
	\text{RCA}_{gs}=\frac{w_{gs}}{\sum_{s'}w_{gs'}}\Big/\frac{\sum_{g'}w_{g's}}{\sum_{g's'}w_{g's'}}
\end{equation}
We thus build the SBN using the binary filter $M_{gs}=1$ if $\text{RCA}_{gs}\ge1$ and $M_{gs}=0$ otherwise. Note that before implementing the filter we apply an exponential smoothing to the RCA series, considering a half-life of 3 years in order to keep a short persistence in the data.

At last we feed to SBN to the Fitness and complexity algorithm~\cite{Tacchella2012,Cristelli2013,Pugliese2016}. 
The method exploits the nested structure of the network and obtains the Fitness $F$ or competitiveness of a geographic area $g$ by aggregating the complexities of its basket of research sectors in a non-linear way (so that the most complex sectors of activity weigh the most), and in the same way the the complexity $C$ of a research sector $s$ is given by the Fitness of the geographic areas that are active in it (with low competitive regions weighting the most).
Operationally, the Fitness and the complexity vectors are the fixed point of the following non-linear iterative map
\begin{eqnarray}
	&\widetilde{F}_g^{(n)}=\sum_s M_{gs} Q_s^{(n-1)}& F_g^{(n)}=\dfrac{\widetilde{F}_g^{(n)}}{\langle\widetilde{F}_g^{(n)}\rangle_g} \nonumber\\ %
	&\widetilde{Q}_s^{(n)}=\dfrac{1}{\sum_g M_{gs} \dfrac{1}{F_g^{(n)}}} &Q_s^{(n)}=\dfrac{\widetilde{Q}_s^{(n)}}{\langle\widetilde{Q}_s^{(n)}\rangle_s}\nonumber
\end{eqnarray}
where the operator $\langle\cdot\rangle_x$ indicates the arithmetic mean with respect to the possible values assumed by the variable dependent on the set $x$. 
Fixed point values of the Fitness are finally normalized by a reference value, which is taken to be the Fitness of United States at TL1 and that of California at TL2.
Fixed point values of the Fitness are finally normalized by a reference value, which is taken to be the Fitness of United States at TL1 and that of California at TL2 (US06).
The normalization aims to regularize the heterogeneous distribution of Fitness among the years, enhancing the relative strength of the nations instead of a global competitiveness.
Note that we build two kind of Fitness indicators: the Scientific Fitness based on log-citations counts of eq. \eqref{eq:logcitation}, and the document Fitness when log-documents counts is used in its place.

We quantify the degree of scientific inequality within a nation using the Gini coefficient estimated on the dispersion of citation counts among its regions~\cite{Cozza2015}
(in the Supporting Information we consider a version of the Gini coefficient that takes population size into account).
For our purposes, the Gini index can be written as follows:
\begin{equation*}
	G = 1 - \frac{\sum_{i+1}^{n} f(y_i)(S_{i-1}+S_i) }{S_n} 
\end{equation*}
where $S_i = \sum_{j=1}^{i} f(y_j)y_j$,  $S_0=0$,  $f(y_i)$ is the fraction of regions within the same country that has received at least $y_i$ citations, and $y_i<y_j$ whenever $i<j$.

We remark that the OAG database allows obtaining the SBN at different levels of geographic aggregations, ranging from the fine-grained description of individual institutions to the macroscale of regions and nations. 
In this work we focus on the macroscopic scale, in order to compare with previous literature of EFC and Science-of-Science.
Leveraging the OECD Territorial Level Classification~\cite{oecd-geo} we generate the SBN both at the Territorial Level 1 (TL1) of nations (207 countries, following the nowadays world structure) as well as at the Territorial Level 2 (TL2), which includes 577 distinct regions ~\footnote{There are in principle more than 700 regions but for some of them there is no affiliation found.} in 43 countries (some of which are not OECD members).

The expenditure database is based on the available data collected by the OECD on the Gross Expenditure in Research and Developments (GERD) indices~\cite{OECD_HERD}.
The database covers 48 countries, \textit{i.e.} all the OECD members and few other relevant nations for which the data is made available, such as China and Russia.
However, the data' quality depends strongly on national features, and the HERD database implemented in the analysis above is made available for 42 nations (among the OECD members on Colombia does not provide information of the expenditure).
We implement a linear interpolation reconstructing the missing points, 
At TL2, the database follows the same classification implemented by the derivation of the territorial level SBN, edited by the OECD.
However, the reconstruction at the lower scales is interpolated keeping constant the national performances, since the data presents more than $50\%$ of missing entries.

\section*{Data Availability}
The datasets generated and analysed during the current study are available in the \emph{Scientific database} repository, \url{https://efcdata.cref.it/}.

\section*{Acknowledgements}
{A.P. gratefully acknowledge funding received from the Joint Research Centre of Seville (grant number 938036-2019 IT).
}

\appendix
\section{Supporting Information Appendix (SI)}

\subsection{Coverage Soft Science / Hard Science}
\begin{figure*}
	\centering
	\includegraphics[scale=0.1]{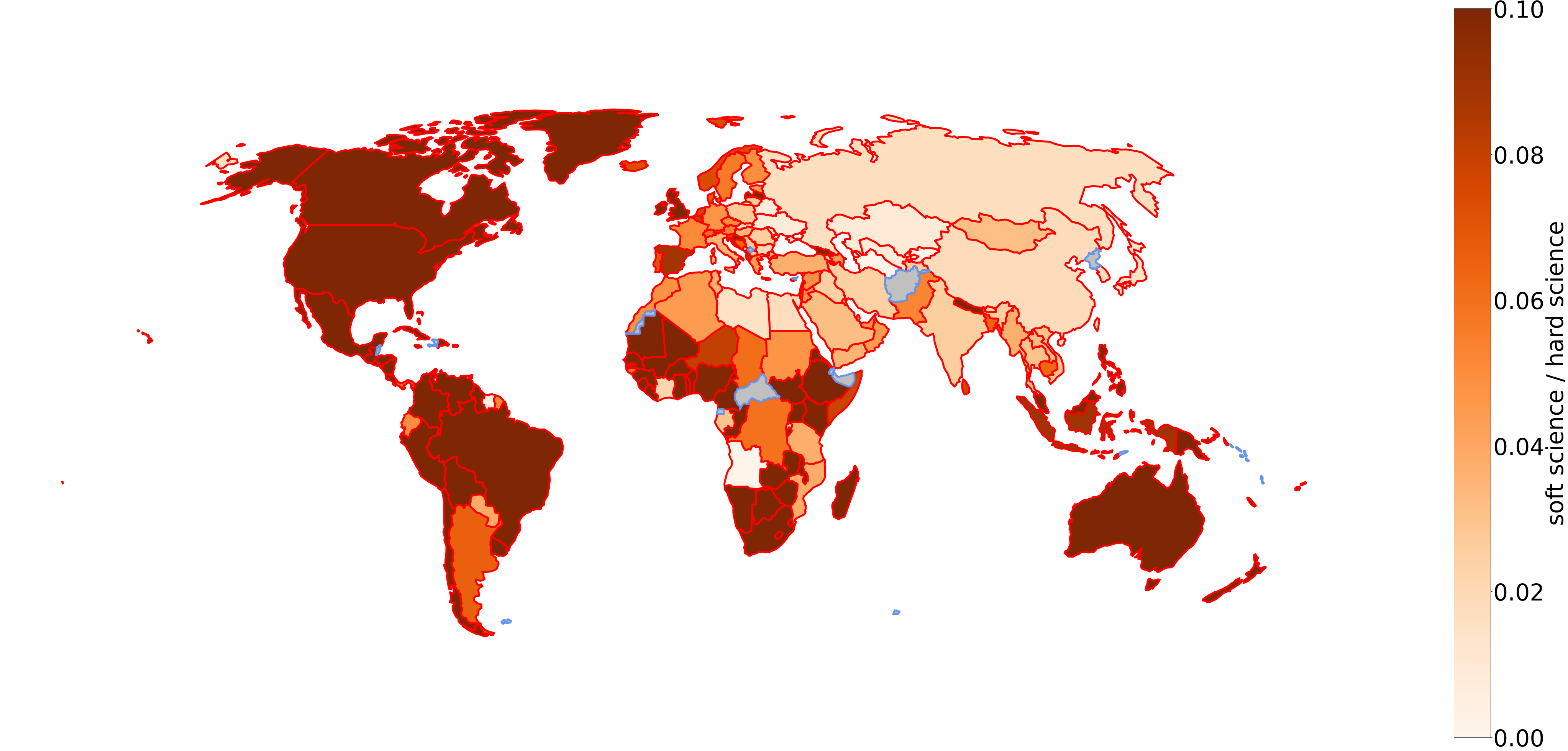}
	\caption{The world map where the color-map correspond to the ration of the overall production of the nations in \textit{soft} science with respect to \textit{hard} science.}
	\label{fig:hardsoft}
\end{figure*}

The Open Academic Graph (OAG) does not provide information on the uniformity of the coverage of the different sectors and nations.
Indeed, a problem faced by others databases, such as SCOPUS, is that English-speaking and developed nations have a full coverage of the literature produced in all the scientific domains, from hard sciences to soft and social sciences, while the rest of the nations may have only a partial coverage, especially in the soft sciences that are mostly written in national languages.
This biases is not addressed by OAG but it can be estimated through the computation of the ratio among the scientific production in hard and soft sciences of the nations.
Defining soft sciences the set of FoS sons of (Sociology, Political Science, Art, Business, Philosophy, History) and hard science the others, OAG presents a language bias that can be visualized in figure~\ref{fig:hardsoft}.

\subsection{Scientific Fitness --- Economic Fitness}
\begin{figure*}
	\centering
	\includegraphics[scale=0.3]{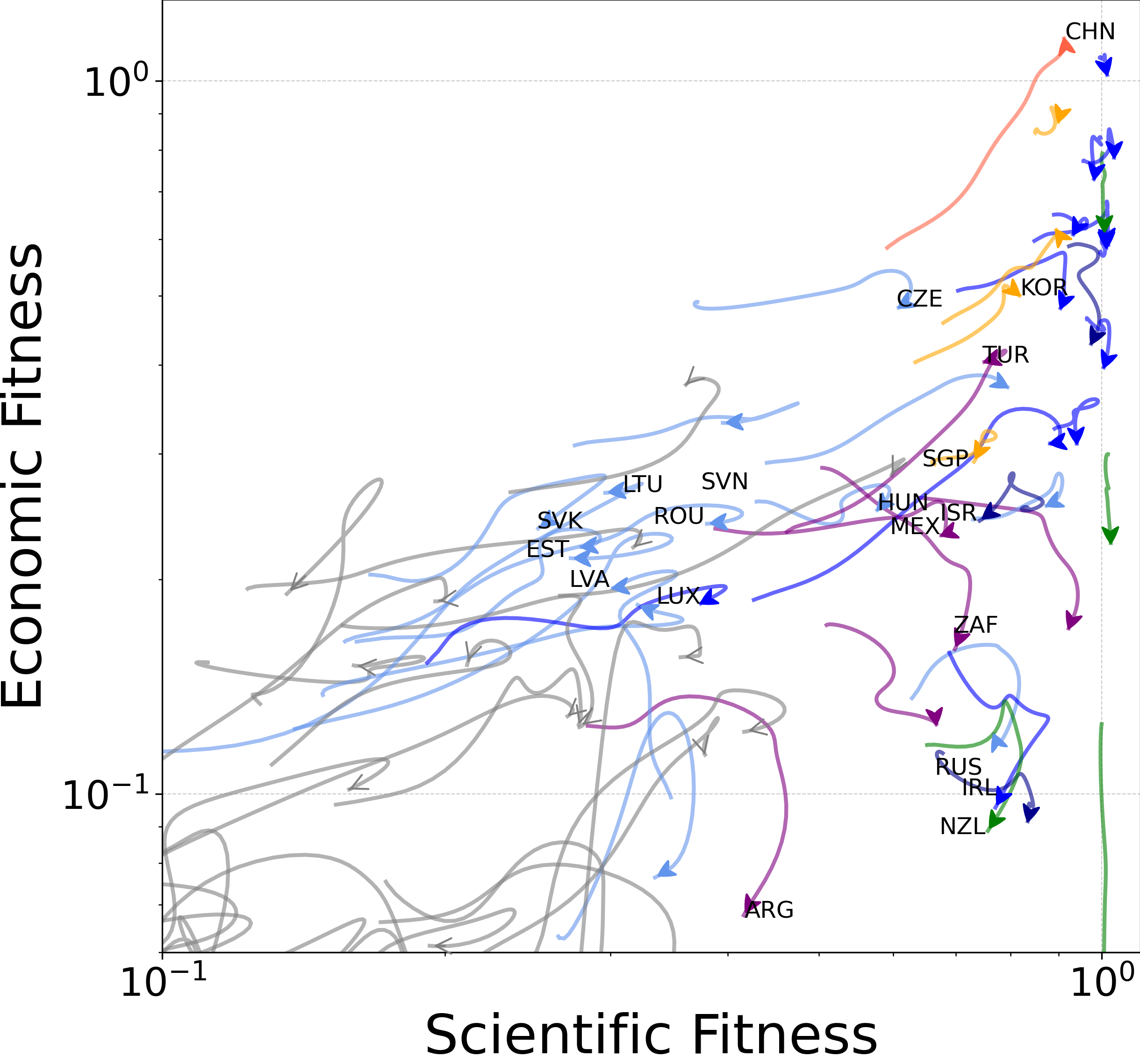}
	\caption{Trajectories of the nations in the diagram scattering the Scientific Fitness and the Economic Fitness}
	\label{fig:scifit-ecofit}
\end{figure*}

The Scientific Fitness is a measure of competitiveness of the national and regional research systems, as discussed in the main text.
The scientific competitiveness depends on the competitiveness on the others sectors of innovations such as the Economic Fitness~\cite{Tacchella2012,JRC_report}, since there is a strong interaction among them.
However, there is no 1-1 relation among the different competitiveness, as shown in figure~\ref{fig:scifit-ecofit}.
Indeed, high Economic Fitness usually translate to high Scientific Fitness while the contrary is not found. 



\subsection{GERD versus Scientific Fitness}
\begin{figure*}
	\centering
	\includegraphics[scale=0.275]{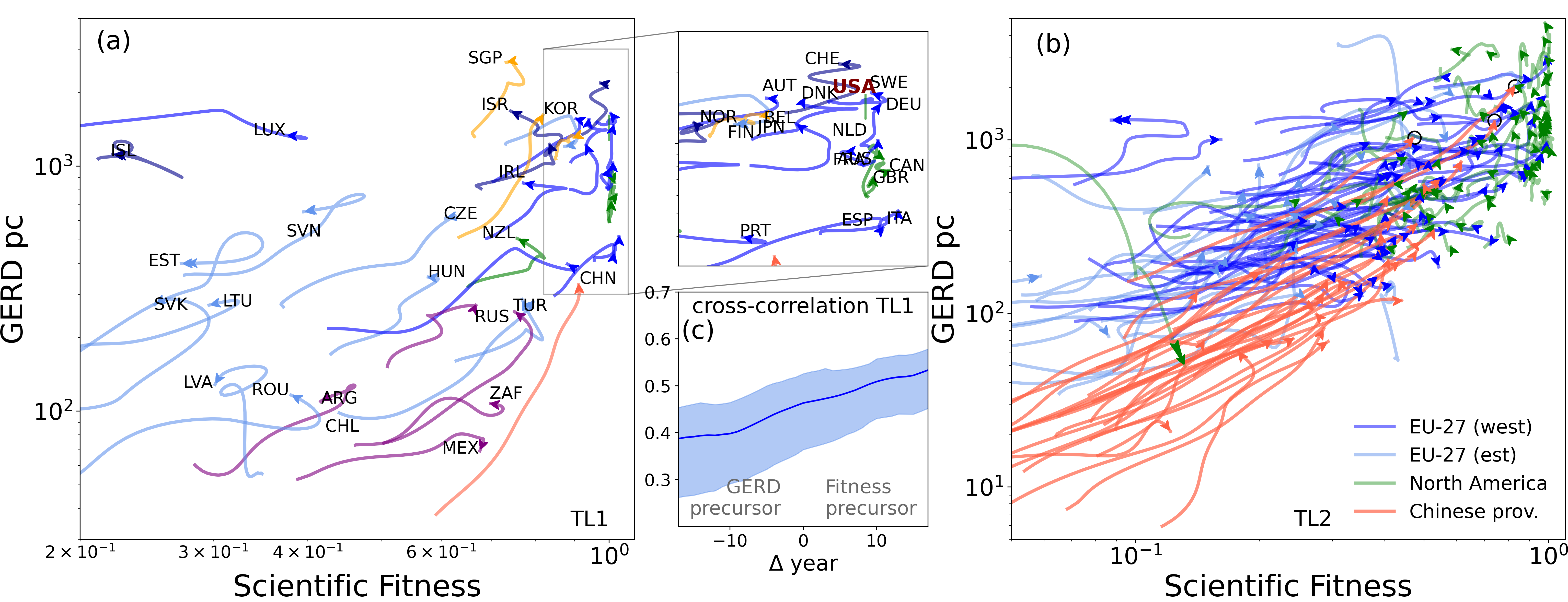}
	\caption{Trajectories of nations (TL1, left panel) in the plane defined by Scientific Fitness and resources invested, the latter  measured by Gross Expenditures on R\&D per capita (GERD-pc). Line colors are used to group nations into macro-areas: dark blue for west EU nations (plus Switzerland, Israel, Norway, Island), light blue for est EU nations, red for China, purple for middle-income countries (Russia, South Africa, Mexico, Argentina, Chile), and green for the English-speaking nations (United States, United Kingdom, Canada, Australia, New Zealand), and yellow for the Asian nations (Singapore, South Korea, Japan). Trajectories represent data from 2000 to 2017, with the arrow indicating the direction of time. 
		The inset zooms on the top-right corner where there is a concentration of highly competitive nations. 
		Trajectories are also displayed for regions (TL2, right panel) belonging to China and a selection of EU west, EU east and North America nations. 
		At last the central panel in the bottom displays the cross-correlation between Scientific Fitness and GERD at the national scale (TL1) averaged over the whole set of countries as a function of the temporal delay ($\Delta$ year) used to compute these quantities. The blue contour represents the $25-75\%$ quantile, generated with a bootstrapping technique. 
	}
	\label{fig:expenditure_diagram_gerd}
\end{figure*}

The scientific expenditure, collected by the OECD, is aggregated in the Gross Expenditure in Research and Developments (GERD), available for 42 nations.
The database can be decomposed in the Governmental expense (GOVERD), the Business part (BERD) and the Higher Educational part (HERD).
Although HERD correlated well with the scientific success, it is possible to derive the same qualitative information considering the gross expenditure, where similar research systems clusters in the expenditure-Fitness diagram.
Figure~\ref{fig:expenditure_diagram_gerd} indicate that the trajectories of the developed nations follow the ones shown in the main text.
Remarkably, the larger difference with the HERD diagram relies on the position of China.
Indeed, gap that China's HERD has with respect to the developed nations can be partially explained by its higher amount of GOVERD expenditures.

\subsection{Inequality metrics with the information of the population size}
\begin{figure*}
	\centering
	\includegraphics[scale=0.3]{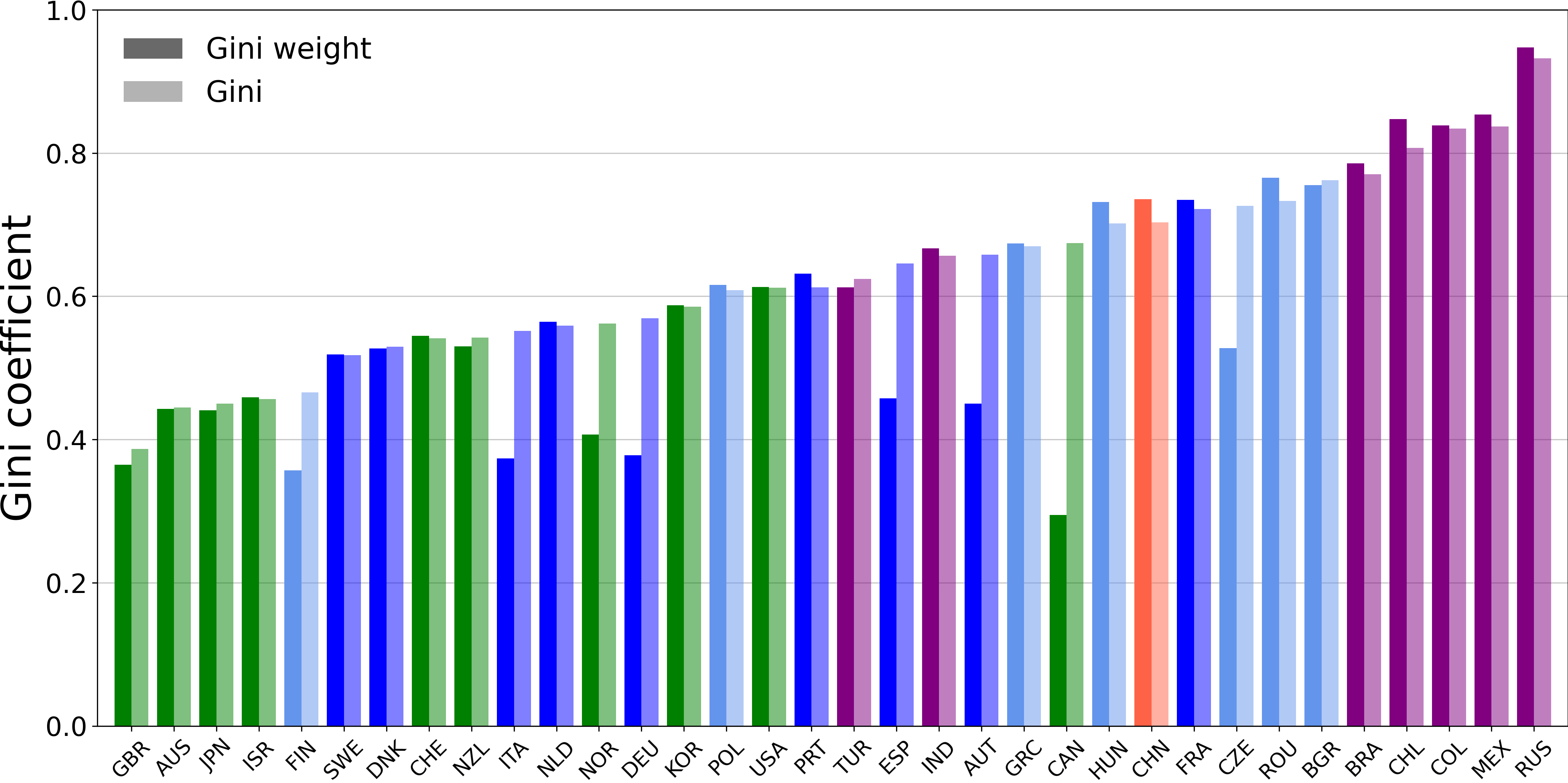}
	\caption{Comparison of the Gini index among the nations computed with and without the estimation of the local number of permanent research as a measure of population size.}
	\label{fig:compar_gini}
\end{figure*}

The inequality implemented in the main text is computed following the procedure discussed in~\cite{Cozza2015}.
However, the computation of the inequality does not account for the different population density and the heterogeneity naturally available on the countries.
Remarkably, the OECD collect the data on the number of permanent researchers~\cite{OECD_HERD} that may be a good estimation of the \textit{population} size in the case of the scientific inequality.
Figure~\ref{fig:compar_gini} shows the bars of the Gini index  as it is implemented in the main text and a weighted version, where the weight are proportional to the sizes of the researcher's population. 

The difference among the measure appears relevant only in some developed nations while it is does not modify the global picture.

\subsection{Comparison of OAG with Scimago}
A second source of data available for the construction of the Scientific Bipartite Network is the database offered by Scimago~\cite{ScimagoDatabase}.
This database is aggregated by Scimagolab using the data available from the SCOPUS~\cite{SCOPUS} database, collected by Elsevier.
The Scimago database offers the matrices of scientific performance (citation counts and document productions among the others) at the national level and implements the full counting statistics: each document assigns a unitary value at each nation having at least an affiliation among the authors.
Thus each national value corresponds to the number of papers produced by researchers operating from the nations, independently on the collaboration sizes.

The scientific classification implemented on the database is the \textit{All Science Journal Classification} (ASJC)~\cite{ASJC}, which gives at the finer level 327 codes, and it is based on the journal classification on which the scientific documents appear.
Thus, the classes does not depend directly from the context of the paper but rather on the topic of the journal, reducing the precision of the analysis based on capabilities.

\begin{figure*}
	\centering
	\includegraphics[scale=0.3]{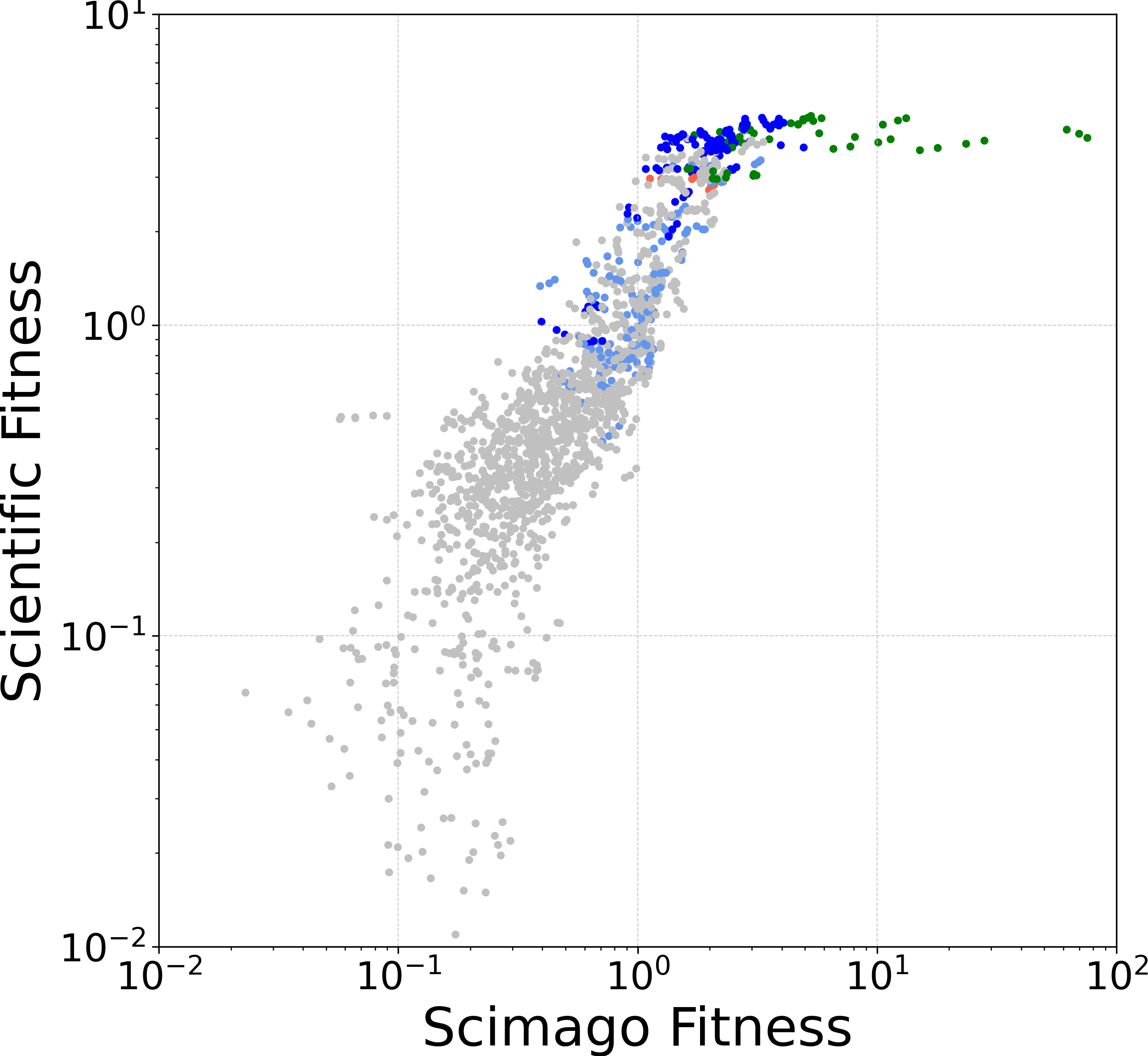}
	\caption{Comparison between the not-normalized Scientific Fitness (using OAG) and the Fitness using Scimago. Each point is a country in a year between $2000$ and $2010$ and the color scheme is: green are the English-speaking nations, blue are the EU nations, red is China. All the others are in gray color.}
	\label{fig:compar_oag_scimago}
\end{figure*}

Despite these aforementioned differences with respect to the OAG database implemented in the main manuscript, the Scientific Fitness computed on the Scimago database does not differs from the one obtained from OAG, except in the subset of the English-speaking nations.
Indeed, the English-speaking nations and primarily the USA outperform the competitiveness of the other nations, collecting most of the global Fitness.
Figure~\ref{fig:compar_oag_scimago} shows the scatter plot of the not normalized Fitness using the SBN based on Scimago and OAG and the English-speaking nations (green dots) is the only set not ling in the main cluster of points.
Removing the outliers, there is a very good correlation between the Fitness based on Scimago and on OAG.
Remarkably, the language bias found in OAG is less dominant with respect to the bias of SCOPUS, in the computation of the Scientific Fitness.

\bibliographystyle{unsrt}
\bibliography{biblio}

\end{document}